\documentstyle[11pt]{article}
\begin{document}
\begin{titlepage}
\vskip0.5cm
\begin{flushright}
CERN TH.7330/94  \\
MS-TPI-94-7 \\
June 1994
\end{flushright}
\vskip0.5cm
\begin{center}
{\Large\bf  Canonical Demon Monte Carlo Renormalization Group}
\end{center}
\vskip 1.3cm
\centerline{M. Hasenbusch$^a$, K. Pinn$^b$, and C. Wieczerkowski$^b$}
\vskip 1.0cm
\centerline{\sl  $^a$ Theory Division, CERN}
\centerline{\sl CH--1211 Geneva 23, Switzerland
\footnote{e--mail: hasenbus~@surya11.cern.ch}}
\vskip .4 cm
\centerline{\sl $^b$ Institut f\"ur Theoretische Physik I,
Universit\"at M\"unster}
\centerline{\sl Wilhelm--Klemm--Str. 9, D--48149 M\"unster, Germany
\footnote{e--mail: pinn/wieczer~@yukawa.uni-muenster.de}}
\vskip 1.cm

\begin{abstract}
\vskip0.2cm
We describe a new method to compute renormalized  coupling constants in
a Monte Carlo renormalization group calculation. The method can be used
for a general class of models, e.g., lattice spin or gauge models. The
basic idea is to simulate a  joint system of block spins and canonical
demons. In contrast to the Microcanonical Renormalization Group invented
by Creutz et al.\ our method does  not suffer from systematical errors
stemming from  a simultaneous use of two different ensembles.
We present numerical results for the $O(3)$ nonlinear $\sigma$-model.
\end{abstract}
\end{titlepage}
%------------------------------------------------------------------------
\newcommand{\nc}{\newcommand}
\nc{\be}{\begin{equation}}
\nc{\ee}{\end{equation}}
\nc{\bea}{\begin{eqnarray}}
\nc{\eea}{\end{eqnarray}}
\nc{\mc}{\multicolumn}
\nc{\half}{\mbox{\small$\frac12$}}
\nc{\quart}{\mbox{\small$\frac14$}}
\nc{\Ra}{$\Longrightarrow$}
\nc{\ra}{$\longrightarrow$}
\nc{\mm}{\hspace{3.3mm}}
%-----------------------------------------------------------------------

\section{Introduction}
The Monte Carlo Renormalization Group (MCRG) \cite{ma} combines  ideas
of the block spin renormalization group (RG) and  Monte Carlo (MC)
simulations.  In the classical MCRG one does not actually compute  the
flow of effective actions (Hamiltonians) or  coupling constants, but
instead uses the RG as a tool to define blocked observables that  are
suitable for the computation of critical properties. Examples for these
techniques are the methods for the determination  of critical exponents
from the linearized RG transformation  \cite{swendsen} and the matching
method for the calculation of the  $\Delta \beta$ function
\cite{wilson}.

However, it would be more in the original spirit  of the RG to really
perform  the integrations over the short wavelength degrees  of freedom
step by step in order to eventually  come close to a fixed point or to
arrive at a correlation length of order one. The
price to pay for this is that one has to deal with a possibly
complicated action with many coupling constants which in addition is not
easy to compute. Furthermore, little is known on the effect of the
truncations  that one has to perform in the number couplings.

In the history of MCRG, some methods have been  invented that allow to
compute renormalized  couplings with the MC method, some restricted  to
specific models, some more generally applicable. For a review on  these
methods see \cite{guptarev}.

Creutz et al.\ have invented a method for the calculation of
renormalized couplings that uses a microcanonical  demon simulation
\cite{creutz_demon}. However, the method  suffers from systematic errors
due to the fact that the  canonical and microcanonical ensemble are
equivalent  only on large lattices.

We here present a modification of the demon method that  overcomes this
disadvantage. Our method does not  introduce systematic errors. It
shares with the Creutz et al.\ method the nice feature that one needs
just a standard  update program for its implementation.

\section{Description of the Method}

We consider a lattice spin system with spins $\phi_x$.
Denote the action  of the spin system by $S$. The Boltzmann factor
is $\exp(-S)$.
Assume that the action can be parameterized by
\be\label{action}
S = - \sum_{\alpha}  \beta_{\alpha} S_{\alpha} \, ,
\ee
where $S_{\alpha}$ are interaction terms, and  the $\beta_{\alpha}$
are real numbers.
In addition we introduce an auxiliary system, called demon system,
that is given by the action
\be
S_D = \sum_{\alpha}  \beta_{\alpha} d_{\alpha} \, ,
\ee
where the $\beta_{\alpha}$ are the same as
in eq.\ (\ref{action}), and the $ d_{\alpha}$
are real numbers in the interval  $[0,d_{max}]$.
In the following we shall consider the joint partition function
\be\label{jointZ}
 Z =
 \left( \prod_{\alpha} \int_{0}^{d_{max}} {\rm d} \, d_{\alpha} \right)
 \int D\phi \, \exp(-S-S_D) \, .
\ee
The partition function factorizes in the partition function of
the spin system and the partition functions of the single demons.
Hence we can compute the distribution of the demon variables $d_{\alpha}$.
One gets
\be
<d_{\alpha}>  = \frac{1}{\beta_{\alpha}}
                \left( 1 -
                \frac{\beta_{\alpha} d_{max}}
                     {\exp(\beta_{\alpha} d_{max})-1}  \right)
\ee
This relation can be numerically solved with respect to
$\beta_{\alpha}$.

Now let us consider the situation of a numerical  RG transformation.  We
simulate the spin model with a known action, and apply a certain
blocking rule to the configurations on the fine grid to produce block
spin configurations on a coarser grid. By this procedure we  get the
blocked configurations with a probability distribution according to
their Boltzmann weight, but without knowing the block effective action
explicitly.

Our proposal how to find the effective action is to perform a simulation
of the joint partition function given in eq.\ (\ref{jointZ}).

We assume that the action of the  blocked spin system is well described
by the ansatz given by eq.\ (\ref{action}). The simulation consists  of
two steps (see figure).

\begin{center}

\begin{tabular}{ccccccccc}
        &          &     &   &   &       &   &            &\\
        &          &     &   &   &       &   &            &\\
        &          &     &   &   &       &   &            &\\
        &          &     &   &   &       &   &            &\\
        &$d$       &     &   &   &$d'$   & = & $d''$      &\\
        &          & \Ra &...&\Ra&       &   &            &\\
        &$\phi$    &     &   &   &$\phi'$&   & $\phi''$   &\\
        &          &     &   &   &$\downarrow$  &   &     &\\
        &          &     &   &   &trash can  &   &        &\\
blocking&$\uparrow$&     &   &   &       &   &$ \uparrow$ &blocking\\
        &          &     &   &   &       &   &            &\\
        &          &     &   &   &       &   &            &\\
        &$\varphi$ & \ra &   &...&       &\ra& $\varphi'$ &\\
\end{tabular}
\end{center}
\begin{center}
\parbox[t]{.85\textwidth}
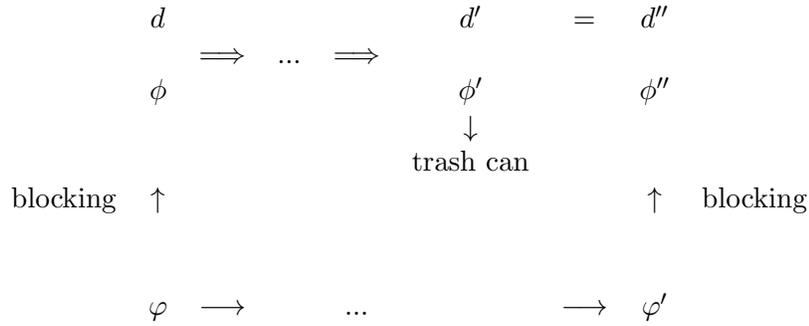
{Figure: Scheme of our procedure to simulate the joint
system of block spins and demons. The symbol \Ra\,\, denotes the
microcanonical updating of block spins $\phi$ and demons $d$,
the symbol \ra\,\, represents the standard updating of the
fine grid spins $\varphi$.
}
\end{center}

\begin{enumerate}
\item
Perform microcanonical updates of the joint system. These updates do not
change the differences $S_{\alpha}-d_{\alpha}$. Since also the Boltzmann
weight is unchanged, knowledge of the $\beta_{\alpha}$ is not required
for the update.

\item
Replace
the block spin configuration by a new, statistically independent one.
\end{enumerate}

The second step is ergodic and fullfils detailed balance if the
simulation of the system on the fine grid satisfies these conditions.
Statistical independence of the block configurations  can be assumed if
the configurations are sufficiently well  separated in computer time,
i.e., if the number of sweeps  between two subsequent configurations is
much larger than  the autocorrelation time of the update algorithm for
the  fine grid spins.

One may ask whether the statistical independence of the  block spin
configurations is really necessary. We studied  an exactly solvable toy
model and found that correlated  block spin configurations deteriorated
the results.

It remains to show that spin-demon update step is ergodic for the
demons. For the concrete update procedure to be described below,  this
property is not too difficult to show.

\section{Numerical Results}

We implemented the new method for the $O(3)$ invariant vector model
in two dimensions.
In particular we considered an action with 12 interaction terms,
\be
S= - \sum_{\alpha=1}^{12} \beta_{\alpha} \sum_x S_{x,\alpha} \, ,
\ee
where
\be
S_{x,\alpha} = \half \sum_{y \in Y_{x,\alpha}} \, (s_x \cdot s_y)^n \, .
\ee

The $s_x$ are 3-vectors of unit length, and the $\cdot$ denotes  the
usual scalar product.  The sets $Y_{x,\alpha}$ consist of all lattice
points that  can be obtained by the obvious symmetry operations from a
representative lattice vector $v$,  and $n$ takes values in $\{1,2,3\}$.
Table 1 gives $n$ and $v$ for $\alpha=1...12$.

\begin{center}
\begin{tabular}{|c|c|c|c|c|c|c|c|c|c|c|c|c|}
\hline
$\alpha$ &1  &2  &3  &4  &5  &6  &7  &8  &9  &10 &11 &12  \\
\hline
$v$      &1,0&1,1&2,0&2,1&3,0&1,0&1,1&2,0&2,1&3,0&1,0&1,1 \\
$n$      &1  &1  &1  &1  &1  &2  &2  &2  &2  &2  &3  &3   \\
\hline
\end{tabular}
 \end{center}
 \begin{center}
  Table 1: Labelling of the 12 interaction terms $S_{x,\alpha}$
\end{center}

As a first test  of the method and the program we tried to reproduce the
couplings on a given lattice. We simulated the model with some ad hoc
chosen set of couplings on an $8^2$ and a $16^2$ lattice. The updates
were performed using a tunable version of the  overrelaxation algorithm.
The demon-spin update can be described as follows. First, a new value
for a single spin is proposed. The probability for the change is
symmetric in the start value and the proposed value.  Then one checks
whether the demons $d_{\alpha}$ can put the changes of the $S_{\alpha}$
in their backpacks, i.e., whether the sums $d_{\alpha} + \Delta
S_{\alpha}$ remain inside the allowed interval $[0,d_{max}]$ for all
$\alpha$. If that is the case the proposal for the spin is accepted, and
the demons are updated to the new value. Otherwise spin and demons  keep
their values.

The spin configurations used for the replacement in the  spin-demon
simulation have to be separated by a number of update steps that is
large compared with the autocorrelation time. In oder to avoid  wasting
too many spin configurations we employed 100 independent  demon systems.
Then we generated a sequence of spin configurations,  where two
successive configurations were separated by a number of sweeps $N_s$
large compared to the autocorrelation time divided by 100.  The
replacement configurations for the 100 demon systems were then
successively taken from this sequence. After a replacement of the spin
configuration, we always performed  one lattice spin-demon updating
sweep.  After a full cycle through the demon systems the measured demon
values were averaged over the 100 systems and written to disk.

\begin{center}
\begin{tabular}{|c|l|l|l|l|l|}
\hline
 & \mc{1}{|c}{couplings}
 & \mc{1}{|c}{$L=8$}
 & \mc{1}{|c}{$L=8$, trunc}
 & \mc{1}{|c}{$L=16$}
 & \mc{1}{|c|}{$L=16$, trunc}     \\
\hline
$\beta_1$ &\mm 1.30  &\mm 1.3010(23) &\mm 1.1399(10) &\mm 1.2999(12) & \mm
1.1408(7) \\
$\beta_2$ &\mm 0.35  &\mm 0.3481(12) &\mm 0.2993(6)  &\mm 0.3485(8)  & \mm
0.3014(4) \\
$\beta_3$ &\mm 0.01  &\mm 0.0105(9)  &--  0.0020(6)  &\mm 0.0098(6)  &  --
0.0017(4) \\
$\beta_4$ &\mm 0.02  &\mm 0.0190(7)  &\mm 0.0138(3)  &\mm 0.0192(4)  & \mm
0.0149(2) \\
$\beta_5$ &\mm 0.004 &\mm 0.0045(6)  &\mm 0.0019(4)  &\mm 0.0039(5)  & \mm
0.0022(2) \\
$\beta_6$ &-- 0.200  &-- 0.2045(30)  &               & -- 0.2002(18) &
  \\
$\beta_7$ &-- 0.080  &-- 0.0808(21)  &               & -- 0.0808(12) &
  \\
$\beta_8$ &-- 0.020  &-- 0.0201(11)  &               & -- 0.0188(6)  &
  \\
$\beta_9$ &-- 0.01   &-- 0.0086(6)   &               & -- 0.0085(4)  &
  \\
$\beta_{10}$ &-- 0.005  &-- 0.0045(8)   &               & -- 0.0038(6)  &
     \\
$\beta_{11}$ &\mm 0.02  &\mm 0.0248(26) &               &\mm 0.0220(14) &
     \\
$\beta_{12}$ &\mm 0.01  &\mm 0.0131(19) &               &\mm 0.0114(12) &
     \\
\hline
$\xi$    &
 \mc{1}{|c|}{57.8(4)} &        &       &       &
 \mc{1}{|c|}{36.7(2)}  \\
 \hline
 \end{tabular}
 \end{center}
 \begin{center}
\parbox[t]{.85\textwidth}
          {Table 2: Reproduction of the original coupling constants. The
           columns labelled by ``trunc'' give the result from a
           truncation to an action with 5 couplings only.}
\end{center}

Our results are presented in Table 2. In addition to the reproduction of
the 12 original couplings we  studied the truncation to a subset of 5
couplings. The truncation experiment served for us as preparation for
the renormalization group study, where one is always  faced with the
necessity to truncate the ansatz for the effective action.

For the $L=8$ lattice we performed $20000$ updates of the 100 demon
systems, with $N_s=1$. In the $L=16$ case we made $10000$ updates, also
with $N_s=1$.  The results show that the method works nicely. Most of
the couplings  are reproduced within 1 sigma error bars.

The last line gives correlation length estimates for the original set
of 12 couplings and also for the couplings obtained from the  truncated
ansatz. The correlation length estimates were computed  using a single
cluster algorithm on a lattice very large compared  to the correlation
length.  The correlation length of the truncated coupling set is
significantly smaller than the original one. A naive truncation of the
original set of couplings (just throwing away the couplings $\beta_6$ to
 $\beta_{12}$) gives a correlation length  larger than 200. On an $L=400$
squared lattice we measured $\xi=195.5 \pm 2.3$.

Next we studied renormalization group transformations, starting from
the standard action with nearest neighbour coupling only.
We used a blocking rule that we call ``dressed decimation''.
All lattice points that have
coordinates $x=(i,j)$ with $i$ {\em and} $j$ even are identified
with block sites.  The block spin $s'$ at site $x$ is then defined
as
\be
s'_x =\frac{ w s_{x} + \quart (1-w) \sum_{y .nn. x} s_y  }
           { | w s_{x} + \quart (1-w) \sum_{y .nn. x} s_y | } \, ,
\ee
where the sum is over the nearest neighbours of $x$.
Tests with the massless free field theory revealed
that the choice $w=0.8$ is a good one in the sense that
the effective actions and especially the fixed point action
had good locality properties.

As a first test we started with $\beta_1=1.9$, and all the other
couplings put to zero. The correlation length for this coupling is
121.2(6) \cite{ullixi}.  We applied the blocking rule described above to
generate the  block configuration used in the replacement step for the
demons.  We always blocked a $32^2$ lattice down to $16^2$ lattice.  The
effective couplings within the truncation scheme of  the 12 couplings
given above were then determined from the  demon expectation values. The
effective couplings were then  used as input for the next iteration
step. Our results for  the first 4 steps are presented in Table 3.  In
the last line we give again correlation length estimates  for the
effective theories, again obtained with a single cluster algorithm on
huge lattices.

Note that in case of an exact renormalization transformation the
correlation length should change exactly by the scale factor of the blocking
rule, that in our case is two. The strong deviations that we observe
clearly indicate that  one should include more couplings in the ansatz
for the effective action.

Note also that there is a very good decay of the couplings with
given $n$ with increasing distance of the spins. We conclude from this and the
systematic errors in the correlation lengths
that probably local interactions with
higher $n$ and also interactions with more than two spins
cannot be left out.
(E.g., 4-point operators proved to be important in the study
of Hasenfratz et al. \cite{hasenfratz} using the classical
approximation.)
Careful studies of these issues, also using alternative methods for the
determination of the effective couplings  and other blocking rules are
underway \cite{tocome}.

\begin{center}
\begin{tabular}{|c|c|l|l|l|l|}
\hline
  \mc{1}{|c}{RG step}
  & \mc{1}{|c}{ 0}
  & \mc{1}{|c}{ 1}
  & \mc{1}{|c}{ 2}
  & \mc{1}{|c}{ 3}
  & \mc{1}{|c|}{ 4}  \\
\hline
 $\beta_1$   &1.9&\mm 1.4619(15)&\mm 1.1764(11)&\mm 0.9785(16)&\mm 0.8201(14)\\
 $\beta_2$   &0.0&\mm 0.2691(9) &\mm 0.3117(8) &\mm 0.2892(8) &\mm 0.2528(8) \\
 $\beta_3$   &0.0& -- 0.0123(6) &\mm 0.0107(5) &\mm 0.0226(6) &\mm 0.0272(4) \\
 $\beta_4$   &0.0&\mm 0.0074(4) &\mm 0.0132(3) &\mm 0.0182(3) &\mm 0.0195(4) \\
 $\beta_5$   &0.0&\mm 0.0043(5) &\mm 0.0031(4) &\mm 0.0032(5) &\mm 0.0034(4) \\
 $\beta_6$   &0.0& -- 0.2213(19)& -- 0.2611(13)& -- 0.2257(14)& -- 0.1689(12)\\
 $\beta_7$   &0.0& -- 0.0800(12)& -- 0.1027(10)& -- 0.0985(13)& -- 0.0766(9)\\
 $\beta_8$   &0.0&\mm 0.0003(9) & -- 0.0057(6) & -- 0.0104(7) & -- 0.0107(8)\\
 $\beta_9$   &0.0& -- 0.0022(5) & -- 0.0011(5) & -- 0.0077(5) & -- 0.0081(5)\\
 $\beta_{10}$&0.0& -- 0.0005(6) & -- 0.0011(5) & -- 0.0014(7) & -- 0.0015(7)\\
 $\beta_{11}$&0.0&\mm 0.0659(17)&\mm 0.0898(13)&\mm 0.0748(18)&\mm 0.0470(15)\\
 $\beta_{12}$&0.0&\mm 0.0327(13)&\mm 0.0440(11)&\mm 0.0425(15)&\mm 0.0301(12)\\
\hline
 $\xi$ & 121.2(6)  &
 \mc{1}{|c}{57.0(4)} &
 \mc{1}{|c}{24.5(2)} &
 \mc{1}{|c}{10.47(5)} &
\mc{1}{|c|}{4.78(2)}  \\
\hline
 \end{tabular}
 \end{center}
\begin{center}
\parbox[t]{.85\textwidth}
{Table 3: Results for the first four RG steps, starting
          from the standard action with $\beta_1=1.9$. The
          last line gives the correlation lengths for
          the effective theories. The successive columns
          give the results of the subsequent RG steps.}
\end{center}

\end{document}